\documentclass[superscriptaddress,prx,twocolumn]{revtex4-1}
\usepackage{graphicx}
\usepackage{amsmath,amssymb,physics,dsfont}
\usepackage{filecontents}
\usepackage{color}
\usepackage{hyperref}
\usepackage{microtype}
\usepackage{braket}
\usepackage{amsmath}
\makeatletter
\newcommand*{\rom}[1]{\expandafter\@slowromancap\romannumeral #1@}
\makeatother

\begin{document}

\title{Revealing Fano Resonance in Dirac Materials ZrTe\texorpdfstring{\textsubscript{5}}{5} through Raman Scattering}

\author{Di Cheng}
\thanks{Equal contribution.}
\author{Tao Jiang}
\thanks{Equal contribution.}
\affiliation{Ames National Laboratory, U.S. Department of Energy, Ames, Iowa 50011, USA}

\author{Feng Zhang}
\affiliation{Ames National Laboratory, U.S. Department of Energy, Ames, Iowa 50011, USA}
\affiliation{Department of Physics and Astronomy, Iowa State University, Ames, Iowa 50011, USA}

\author{Genda Gu}
\affiliation{Condensed Matter Physics and Materials Sciences Department, Brookhaven National Laboratory, Upton, NY 11973-5000, USA}

\author{Liang Luo}
\affiliation{Ames National Laboratory, U.S. Department of Energy, Ames, Iowa 50011, USA}

\author{Chuankun Huang}
\author{Boqun Song}
\affiliation{Ames National Laboratory, U.S. Department of Energy, Ames, Iowa 50011, USA}
\affiliation{Department of Physics and Astronomy, Iowa State University, Ames, Iowa 50011, USA}

\author{Martin Mootz}
\affiliation{Ames National Laboratory, U.S. Department of Energy, Ames, Iowa 50011, USA}

\author{Avinash Khatri}
\affiliation{Ames National Laboratory, U.S. Department of Energy, Ames, Iowa 50011, USA}
\affiliation{Department of Physics and Astronomy, Iowa State University, Ames, Iowa 50011, USA}

\author{Joong-Mok Park}
\affiliation{Ames National Laboratory, U.S. Department of Energy, Ames, Iowa 50011, USA}

\author{Qiang Li}
\affiliation{Condensed Matter Physics and Materials Sciences Department, Brookhaven National Laboratory, Upton, NY 11973-5000, USA}
\affiliation{Department of Physics and Astronomy, Stony Brook University, Stony Brook, New York 11794-3800, USA}  

\author{Yongxin Yao}
\affiliation{Ames National Laboratory, U.S. Department of Energy, Ames, Iowa 50011, USA}
\affiliation{Department of Physics and Astronomy, Iowa State University, Ames, Iowa 50011, USA}

\author{Jigang Wang}
\email{jgwang@iastate.edu}
\affiliation{Ames National Laboratory, U.S. Department of Energy, Ames, Iowa 50011, USA}
\affiliation{Department of Physics and Astronomy, Iowa State University, Ames, Iowa 50011, USA}

\begin{abstract}
We explore the Fano resonance in ZrTe\textsubscript{5}, using Raman scattering measurements. We identified two closely spaced B\textsubscript{2g} phonon modes, B\textsubscript{2g} \rom{1} and B\textsubscript{2g} \rom{2}, around 9 meV and 10 meV, respectively. Interestingly, only B\textsubscript{2g} \rom{1} exhibited the Fano resonance, an outcome of quantum interference between discrete phonon modes and continuous electronic excitations. This is consistent with the much stronger electron-phonon coupling of B\textsubscript{2g} \rom{1} mode demonstrated by first-principles calculations. 
Additionally, temperature-dependent measurements highlight an enhanced Fano asymmetry at elevated temperatures, contributed by the thermal renormalization of the band structure and electron-phonon coupling.
This study offers insights into the complex interrelation of electron-phonon coupling, thermal effects, and Fano resonances in ZrTe\textsubscript{5}.
\end{abstract}
\maketitle

\begin{figure*}[t]
\includegraphics[clip,width=6.5in]{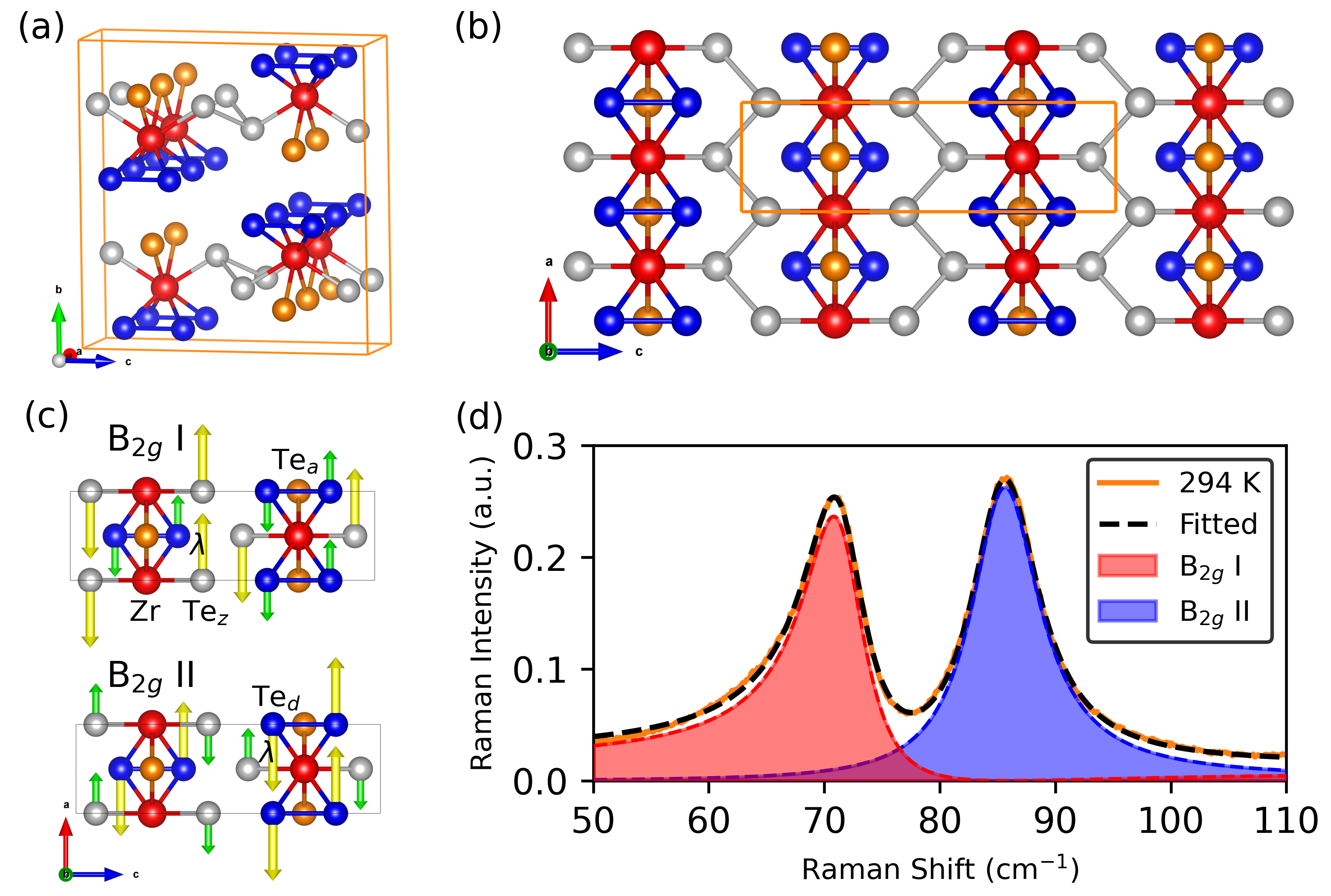}
\caption{Crystal structure of ZrTe\textsubscript{5} viewed from (a) a axis and (b) b axis. The red spheres represent Zr atoms. The orange spheres represent apical Te (Te\textsubscript{a}). The blue spheres represent dimer Te (Te\textsubscript{d}) atoms. The grey spheres represent the zigzag Te (Te\textsubscript{z}) atoms. (c) Eigenvector of the B\textsubscript{2g} \rom{1} and B\textsubscript{2g} \rom{2} modes viewed from b-axis. Yellow represents long vectors, and green represents short vectors, with the length of the long one being 1.64 times of the short one. We note that the maximum atomic displacement $\lambda$ is the same in both modes of equal amplitude, which is associated with Te\textsubscript{z} in B\textsubscript{2g} \rom{1} and Te\textsubscript{d} in B\textsubscript{2g} \rom{2} mode, respectively.(d) B\textsubscript{2g} \rom{1} and B\textsubscript{2g} \rom{2} modes at room temperature. The orange solid line shows the Raman spectrum measured at 294 K, while the black dashed line represents the fitted curve. The shaded areas, outlined with dashed lines, illustrate the components of the two-peak fit: the red shade corresponds to the B\textsubscript{2g} \rom{1} mode, and the blue shade corresponds to the B\textsubscript{2g} \rom{2} mode.}
\label{Fig1}
\end{figure*}

\section{Introduction}
The intricate realm of quantum interference between electronic and phononic excitations has consistently been a fertile ground for novel discoveries in condensed matter physics\cite{Wipf1978, ugeda2016,ishigami2007,peterfalvi2022quantum}. A manifestation of such interference is the Fano resonance. First elucidated by Ugo Fano in 1961\cite{fano1961effects}, this resonance emerges due to the interference between a discrete state and a continuum of states, resulting in asymmetric spectral line shapes. Over the years, Fano resonance has garnered considerable attention, not just for its foundational significance in quantum mechanics, but also for its pivotal role in diverse fields ranging from photonics and optoelectronics to biosensing and solar cells\cite{gallinet2011ab,rahmani2012subgroup,hao2008symmetry,giannini2011plasmonic}.

Dirac semimetals are characterized by the linear dispersion of electronic bands that touch at discrete points in the Brillouin zone\cite{xu2015discovery,bradlyn2016beyond}. 
ZrTe\textsubscript{5}, a representative of this class, has garnered significant attention in recent years\cite{kim2013dirac,zhang2017electronic,li2016chiral,kim2021terahertz,kim2023sub,jiang2023ab, haeuser2024analysis}. Its unique crystal structure and inherent symmetries have made it a paradigm for investigating the phenomena associated with Dirac fermions\cite{tabert2016optical}. Not only does ZrTe\textsubscript{5} exhibit a strong anisotropy in its electronic and thermal transport properties, but its behavior under external stimuli, such as magnetic fields, showcases exotic phenomena that challenge our conventional understanding of solid-state physics\cite{chen2015,zhu2015quantum}. Recently, the phonon mode-selective symmetry switching effects has been discovered in ZrTe\textsubscript{5} and provide compelling implications of dynamical topological control with coherent phonon pumping with selective IR\cite{luo2021light} and Raman\cite{vaswani2020light} symmetries. The importance of ZrTe\textsubscript{5} goes beyond its fundamental interest. Its unique properties open doors to potential applications in high-speed electronics, spintronics, and quantum computing\cite{schoop2016dirac,liu2014tuning}.This has triggered studies of dynamic stability and coherent phonons in topological materials in both experiment\cite{cheng2024chirality,luo2021ultrafast,yang2020light,luo2019ultrafast,luo2025symmetry,luo2023room,kim2021terahertz} and theory\cite{song2025quantum,song2024quantum}.

Notably, while the importance of Fano resonance is well-recognized, few studies have systematically approached its implications in specific materials such as Dirac semimetals\cite{xu2017temperature}. This paper delves into an in-depth investigation of the Fano resonance observed in Raman scattering measurements on the sample of ZrTe\textsubscript{5}. Our exploration of ZrTe\textsubscript{5} uncovers two B\textsubscript{2g} phonon modes in proximity, i.e., B\textsubscript{2g} \rom{1} and B\textsubscript{2g} \rom{2}, situated around the 10 meV frequency region. Intriguingly, while these modes are closely spaced in terms of frequency, only B\textsubscript{2g} \rom{1} exhibits the characteristic Fano resonance. Such a distinction raises a pivotal question: What underlying factors influence this differential behavior between two seemingly similar phonon modes?

To address this, it's essential to understand the essence of the Fano resonance. Rooted in the quantum interference phenomena, the resonance pattern is an outcome of quantum interference between discrete phonon modes and continuous electronic excitations, which ties to the electron-phonon coupling effects. By performing first-principles frozen-phonon calculations for both B\textsubscript{2g} Raman modes, we show that the band structure, in particular band gap renormalization, is much more sensitive to the lattice displacement of B\textsubscript{2g} \rom{1} phonon, which supports the strong mode-selective electron-phonon coupling and hence the observed disparity in Fano resonance between the two modes.

Complementing our primary findings, we also embarked on temperature-dependent Raman measurements. The insights garnered from this facet of our study accentuate the Fano asymmetry's enhancement with increasing temperatures. This observation cannot be explained by the pure thermal smearing effect for the band occupations, where the Fano asymmetry is expected to become weaker with increasing temperature due to thermal blocking effect for the electronic state transition~\cite{xu2017temperature}. With first-principles calculations, we show that the peculiar thermal enhancement for Fano resonance originates from the thermal renormalization effect for the band structures of the ZrTe\textsubscript{5} system, manifested by the experimentally observed temperature-induced Lifshitz transition~\cite{zhang2017electronic}. 

In the subsequent sections, we systematically detail our experimental setup, present our findings, and offer a comprehensive analysis that connects electron-phonon coupling, thermal effects and the intriguing Fano resonance in the context of ZrTe\textsubscript{5}.

\section{Material and methods}
The growth of the ZrTe\textsubscript{5} single crystal was accomplished using the flux growth method, where Te acted as the flux. We employed high-purity elements: 99.99999\% Te and 99.9999\% Zr. These elements were placed inside a double-walled quartz ampoule and sealed in a vacuum. The melt composition for crystal growth was Zr\textsubscript{0.0025}Te\textsubscript{0.9975}. The mixture was first melted at 900 \textdegree C in a box furnace and stirred continuously for 72 hours to achieve homogeneity. This was followed by a controlled cooling, then a quick reheating, aiming to re-melt microcrystals in a 445-505 \textdegree C range over 21 days. The structural details of ZrTe\textsubscript{5} can be seen in Figs.\ \ref{Fig1}(a) and \ref{Fig1}(b).

For the Raman spectra collection, a 784.5 nm continuous laser served as the excitation source. It passed through four bandpass filters (BPF) to enhance spectral clarity. The setup was arranged in a backscattering unpolarized geometry with the sample placed inside a liquid-helium-cooled cryostat. Three Notch filters (BNF) were used to block the Rayleigh scattering. A monochromator paired with a Charge-Coupled Device (CCD) detector was used for signal detection. The power of the laser on the sample is 4 mW, and the focused diameter is 10 $\mu$m.

\section{Results and discussion}

In Figure \ref{Fig1}(c), we present the atomic displacements linked with the B\textsubscript{2g} modes. The corresponding room-temperature Raman spectrum (294 K), showing the B\textsubscript{2g} \rom{1} and B\textsubscript{2g} \rom{2} modes, is presented in Fig.\ \ref{Fig1}(d). The experimental Raman spectrum at 294 K is shown as an orange solid line, and the black dashed line denotes the fitted curve. A two-peak fitting model is applied, where the shaded areas outlined by dashed lines correspond to the individual phonon modes: the red shade denotes the B\textsubscript{2g} \rom{1} mode, and the blue shade denotes the B\textsubscript{2g} \rom{2} mode. The asymmetric lineshape of the B\textsubscript{2g} \rom{1} peak stands in contrast to the symmetric Lorentzian profile of B\textsubscript{2g} \rom{2}, clearly illustrating the Fano resonance in the former.

\begin{figure*}[t]
\includegraphics[clip,width=6.5in]{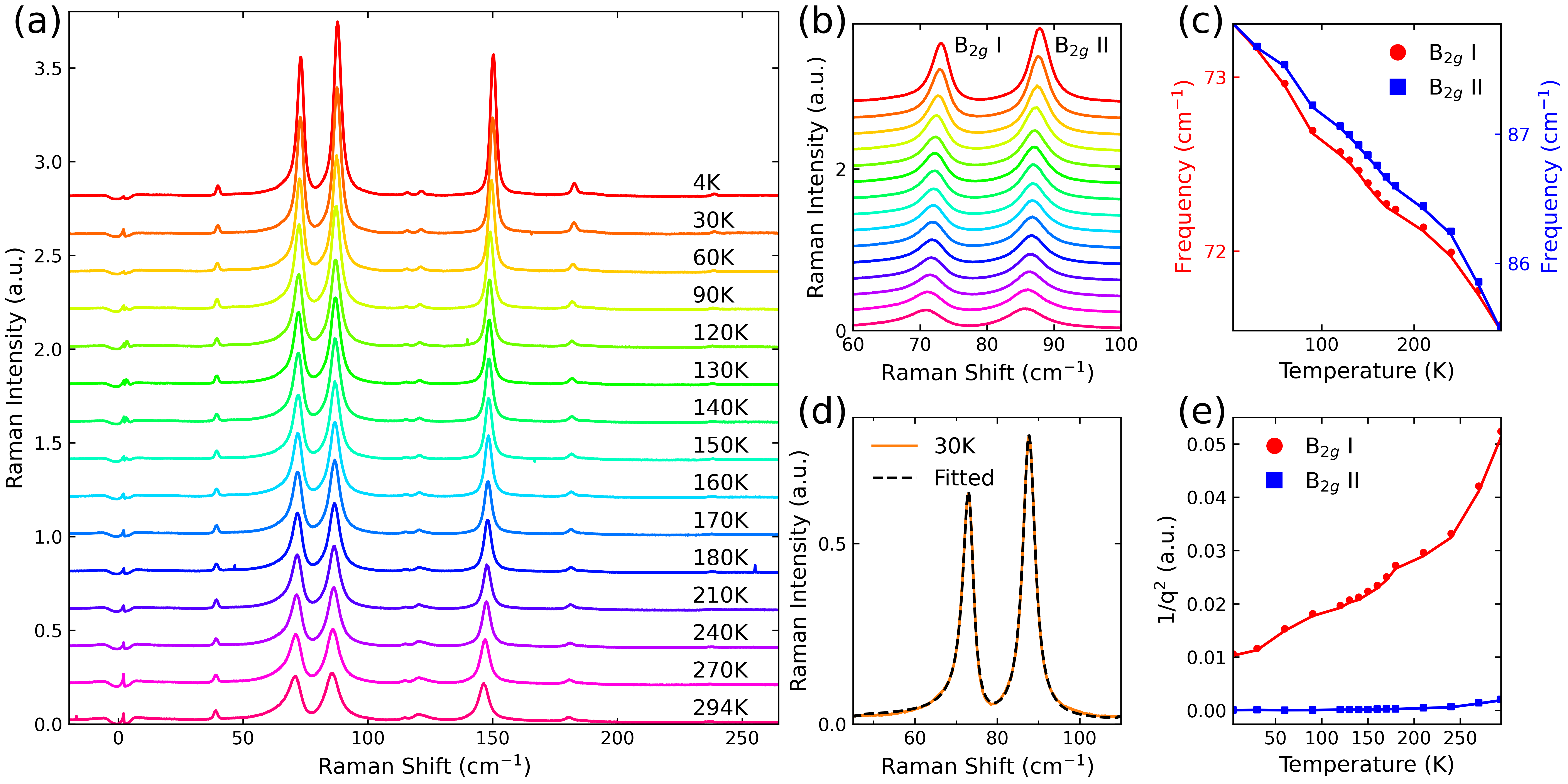}
\caption{Raman spectra of ZrTe\textsubscript{5} as a function of temperature: (a) ranging from -20 to 260 cm\textsuperscript{-1}; (b) spanning 60 to 100 cm\textsuperscript{-1}. (c) Frequency variation with temperature for B\textsubscript{2g} \rom{1} mode (red dots) and B\textsubscript{2g} \rom{2} mode (blue squares) with error bars. (d) Comparison between original (orange) and fitted (dashed black) line shapes of the two B\textsubscript{2g} Raman modes at 30 K, with fitting parameters detailed in equation (2). (e) Temperature-dependent $1/q^2$ for the B\textsubscript{2g} \rom{1} (red dots) and B\textsubscript{2g} \rom{2} (blue squares) Raman modes with error bars.}
\label{Fig2}
\end{figure*}

A comprehensive temperature-dependent evolution of these peak shapes is displayed in Fig.\ \ref{Fig2}(a), extending the Raman spectra measurements to 4 K. Notably, both peaks become sharper at reduced temperatures. To emphasize the temperature-dependent behavior of these two peaks, Fig.\ \ref{Fig2}(b) provides a focused view. As shown in Fig.\ \ref{Fig2}(c), both peaks exhibit a linear decrease in frequency with rising temperature despite their initial frequency differences. A more detailed analysis of the phonon frequency shifts using the anharmonic decay model is presented in the Supplementary Information, further confirming that anharmonic effects dominate within the studied temperature range.
The observed asymmetric profile of the B\textsubscript{2g} \rom{1} mode is typically referred to as the Fano line shape and can be fitted by the following equation\cite{zhang2011raman}:
\begin{equation}
    I(\omega)=I\textsubscript{0}\frac{[1+(\omega-\omega\textsubscript{0})/q\Gamma)]\textsuperscript{2}}{1+[(\omega-\omega\textsubscript{0})/\Gamma]\textsuperscript{2}}
\end{equation}

where $I\textsubscript{0}$ is the intensity, $\omega\textsubscript{0}$ is the phonon frequency in the context of Fano resonance, $\Gamma$ controls the line width, and q is an asymmetry indicator with a smaller $|q|$ resulting in pronounced asymmetry. In the specific scenario where $q \to \infty$, the equation simplifies to the conventional Lorentzian shape. For an accurate characterization, we applied this equation to fit the Raman spectra of B$_{2g}$ \rom{1} mode at varying temperatures. Owing to the slight overlap between the two B\textsubscript{2g} modes, B$_{2g}$ \rom{2} mode, represented by a Lorentzian shape, was incorporated in the fitting. More precisely, the Raman spectrum within the 49 cm\textsuperscript{-1} and 110 cm\textsuperscript{-1} range was matched to the subsequent equation:
\begin{equation}
    I(\omega)=I\textsubscript{1}\frac{[1+(\omega-\omega\textsubscript{1})/q\Gamma\textsubscript{1})]\textsuperscript{2}}{1+[\omega-\omega\textsubscript{1}/\Gamma\textsubscript{1}]\textsuperscript{2}}+I\textsubscript{2}\frac{1}{1+[(\omega-\omega\textsubscript{2})/\Gamma\textsubscript{2}]\textsuperscript{2}}+C
\end{equation}

Wherein, the subscripts 1 and 2 respectively represent B$_{2g}$ \rom{1} and B$_{2g}$ \rom{2} modes. An additional parameter C accounts for the background contribution, primarily from the CCD camera's dark current. A detailed justification for using a constant background parameter in the Fano lineshape fitting is provided in the Supplementary Information. Illustratively, Fig.\ \ref{Fig2}(d) offers fitting outcomes at T=30 K, reinforcing the efficiency of our approach as evidenced by the alignment between the fitted and observed values. Lastly, in Figure \ref{Fig2}(e), we illustrate the temperature dependency of the parameter $1/q^2$ for B$_{2g}$ \rom{1} mode. The consistent increase in $1/q^2$ underscores the positive correlation between Fano asymmetry and temperature.

\begin{figure*}[t]
\includegraphics[clip,width=6.5in]{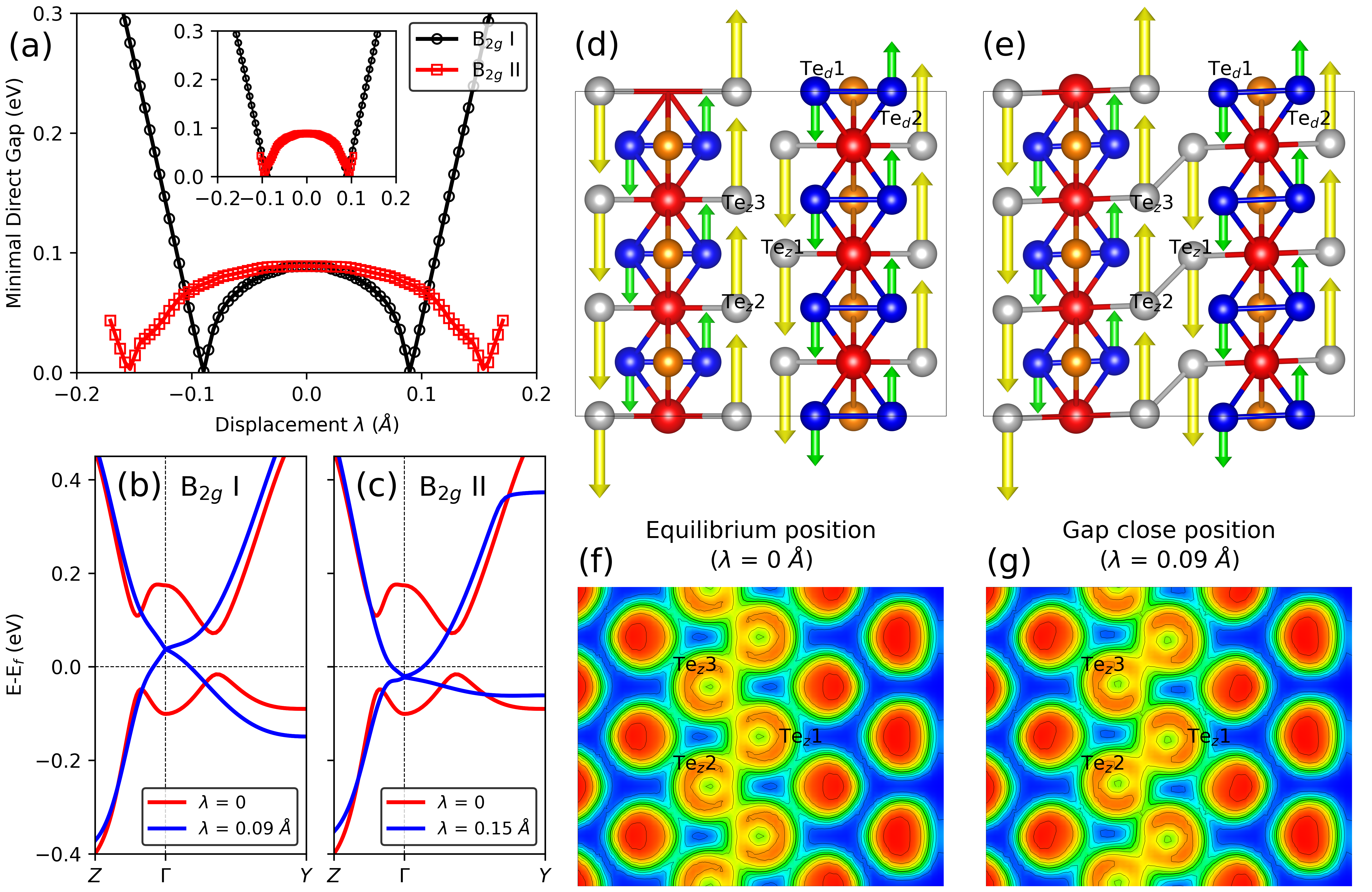}
\caption{(a) Minimal band gap as a function of atomic displacement of Te\textsubscript{z} for B\textsubscript{2g} \rom{1} mode (black circle) and Te\textsubscript{d} for B\textsubscript{2g} \rom{2} mode (red squares). Inset shows the minimal band gap as a function of Te\textsubscript{z} displacement for both B\textsubscript{2g} \rom{1} mode (black dots) and B\textsubscript{2g} \rom{2} mode (red squares). (b) Band structures for B$_{2g}$ \rom{1} mode with displacement $\lambda$=0 (red) and $\lambda$=0.09 \text{\AA} (blue). (c) Band structures for B$_{2g}$ \rom{2} mode with displacement $\lambda$=0 (red) and $\lambda$=0.15 \text{\AA} (blue). (d, e) Crystal structure of ZrTe\textsubscript{5} at ac plane with B\textsubscript{2g} \rom{1} mode displacement at $\lambda$=0 (Equilibrium position) and $\lambda$=0.09 \text{\AA} (gap close position), highlighting the reduced Te\textsubscript{z1}-Te\textsubscript{z2} distance. The B\textsubscript{2g} \rom{1} phonon eigenvector is also shown from reference. (f, g) The corresponding 2D contour plot of electron localization function.}
\label{Fig3}
\end{figure*}

\section{Theoretical calculations and analysis}
Theoretically, the dimensionless parameter $q$ characterising Fano asymmetry can be evaluated according to Ref~\cite{fano1961effects},Ref~\cite{tang2010tunable}, and Ref~\cite{xu2017temperature}
\begin{equation}
q = \frac{1}{\pi V_\textrm{e-ph} D_\textrm{e-h}(\omega_0, T)} \times \frac{\mu_\textrm{ph}}{\mu_\textrm{e-h}},
\label{eq: q}
\end{equation}
where $\mu_\textrm{ph}$ and $\mu_\textrm{e-h}$ are the optical matrix elements for dressed phonon and electron-hole pair excitations, respectively. $V_\textrm{e-ph}$ is the electron–phonon coupling strength, and $D_\textrm{e-h}(\omega, T)$ is the temperature-dependent joint electron–hole pair density of states at a given phonon frequency $\omega_0$. While in principle the ratio of the optical matrix elements $\mu_\textrm{ph}/\mu_\textrm{e-h}$ depends on the specific phonon, here we focus on the analysis of electron-phonon coupling effect to explain the selective Fano resonance of B\textsubscript{2g} \rom{1} Raman mode.

To compare the electron-phonon coupling strength between the two B\textsubscript{2g} modes, 
we performed frozen-phonon calculations based on density functional theory (DFT). We employed VASP package\cite{VASP} and adopted the local density approximation (LDA) for the exchange-correlation energy functional~\cite{ceperley1980ground,perdew1981self}. 
When determining the relaxed lattice parameters for ZrTe$_5$, we set a $10\times 3\times 3$ $\Gamma$-centered uniform $\mathbf{k}$-mesh and a plane-wave cutoff energy of 230 eV for the conventional cell, which contains four formula units. Energy convergence criterion is set to $10^{-9}$ eV for electronic self-consistency and 0.001 eV/\AA\ for ionic relaxation. The calculated lattice constants are a=3.94 \AA, b=14.29 \AA\ and c=13.50 \AA.\ The computed lattice constants match well with experimental data of a=3.99 \AA\, b=14.50 \AA\ and c=13.73 \AA~\cite{wu2016evidence}. 
Electronic structure calculations employed a 10$\times$10$\times$6 k-mesh with the primitive unit cell containing two formula units. All calculations incorporated the spin-orbit coupling effect. 
For calculating the phonon modes at the Brillouin zone center ($\Gamma$ point), we employed the finite displacement method available in the PHONONPY package~\cite{phonopy}. The calculated phonon frequencies for the two B\textsubscript{2g} modes are 73.2 cm$^{-1}$ and 89.4 cm$^{-1}$, which are in close agreement with the experimental measurement of 73.3 cm$^{-1}$ and 87.8 cm$^{-1}$, respectively.

To perform frozen phonon calculations, we introduce the atomic displacements according to the phonon eigenvector. For the equilibrium structure of ZrTe$_5$, the band edge for both the valence and conduction bands deviate from the zone center toward $\Gamma$-Z and $\Gamma$-Y directions, and a direct band gap of 89 meV is located at the $\Gamma$-Y high-symmetry line as shown by red curve in Fig.~\ref{Fig3}(b) and (c). Fig.~\ref{Fig3}(a) illustrates the band gap sizes as a function of atomic displacement $\lambda$ for the two B\textsubscript{2g} modes. Here $\lambda$ is defined as the displacement of Te\textsubscript{z} atom for B\textsubscript{2g} \rom{1} mode and that of Te\textsubscript{d} atom for B\textsubscript{2g} \rom{2} mode, because they represent the maximal atomic displacement of the same amplitude in the respective eigenvector (see Fig.~\ref{Fig1}(c)). In other words, $\lambda$ is proportional to the phonon amplitude. 
Clearly, the electronic structure is more sensitive to the lattice modulation of B\textsubscript{2g} \rom{1} mode as the band gap renormalization shows. Therefore, the electron-phonon coupling strength of B\textsubscript{2g} \rom{1} mode is much stronger than that of B\textsubscript{2g} \rom{2} mode, which is consistent with the experimentally observed mode-selective Fano resonance for the B\textsubscript{2g} \rom{1} phonon. We also note that with increasing the phonon amplitude, the band gap initially decreases and turn to increase after reaching zero at $\lambda \approx 0.09$ \AA\ for B\textsubscript{2g} \rom{1} mode and $\lambda \approx 0.15$ \AA\ for B\textsubscript{2g} \rom{2} mode. This is reminiscent of the A\textsubscript{1g} Raman modulated topological phase transition in ZrTe\textsubscript{5}~\cite{vaswani2020light}.

Deeper insight about the disparity of electron-phonon coupling strength can be obtained by a further comparison between the two B\textsubscript{2g} Raman modes. As shown in Fig.~\ref{Fig1}(c), the two modes display a quite similar oscillation pattern along a-axis, with the exception that displacement vectors associated with Te\textsubscript{z} and Te\textsubscript{d} are switched. This difference is nevertheless crucial. First, the displacement magnitude associated Te\textsubscript{z} and Te\textsubscript{d} are significantly different, with a ratio of $1.6$ in the B\textsubscript{2g} \rom{1} mode and $1/1.6\approx 0.62$ in B\textsubscript{2g} \rom{2} mode. Second, electron localization function (ELF) analysis shows that the band gap modulation is tied to the covalency enhancement of the Te\textsubscript{z}-Te\textsubscript{z} bond. Specifically, in Fig.~\ref{Fig3}(f-g) we show the contour plot the ELF of the ZrTe\textsubscript{5} at equilibrium structure ($\lambda=0$) and that at $\lambda=0.09$ \AA\ for the B\textsubscript{2g} \rom{1} mode, where the band gap vanishes. These band structures are shown in Fig.~\ref{Fig3}(b). The bond covalency between Te\textsubscript{z}-1 and Te\textsubscript{z}-2 is clearly strengthened as evidenced by the less steep variation of ELF at the bond center when $\lambda$ increases. The covalent bond enhancement is facilitated by the distance between Te\textsubscript{z}-1 and Te\textsubscript{z}-2 atoms, which reduces from 2.92 \AA\ to 2.80 \AA\ as also shown in Fig.~\ref{Fig3}(d) and (e). The inset of Fig.~\ref{Fig3}(a) further shows a similar variation of the band gap for both modes with respect to the same Te\textsubscript{z} atomic displacement. Equivalently, Fig.~\ref{Fig3}(c) plots the band structure for B\textsubscript{2g} \rom{2} mode with displacement $\lambda=0.15$ \AA\ for the Te\textsubscript{d} (about 0.09 \AA\ for the Te\textsubscript{z}) when the band gap closes. Since B\textsubscript{2g} \rom{1} mode has much larger displacement component of Te\textsubscript{z} atom, it is more effective to modulate bond covalency than B\textsubscript{2g} \rom{2} mode and hence much larger electron-phonon coupling.

While both B\textsubscript{2g} \rom{1} and B\textsubscript{2g} \rom{2} modes exhibit similar linear redshifts in frequency with increasing temperature [Fig.~\ref{Fig2}(c)], consistent with standard anharmonic phonon decay, no additional softening attributable to enhanced electron-phonon coupling is observed for the B\textsubscript{2g} \rom{1} mode. This suggests that, within the studied temperature range, anharmonic lattice effects dominate the phonon frequency renormalization. The stronger electron-phonon coupling associated with the B\textsubscript{2g} \rom{1} mode manifests more prominently in the asymmetric lineshape and linewidth broadening, rather than in its frequency shift.


So far, we have understood the mode-selective Fano resonance through the stronger electron-phonon coupling of the B\textsubscript{2g} \rom{1} mode. We move on to analyze the key physical contributions to the experimentally observed temperature-enhancement of Fano resonance by considering the temperature-dependent joint electron-hole pair density $D_\textrm{e-h}(\omega, T)$ in Eq.~\eqref{eq: q}, which can be evaluated as
\begin{equation}
D_\textrm{e-h}(\omega, T) = \sum_\mathbf{k} w_\mathbf{k}\sum_{i<j} \delta(\omega-(\epsilon_{j\mathbf{k}} - \epsilon_{i\mathbf{k}}))(f(\epsilon_{i\mathbf{k}})-f(\epsilon_{j\mathbf{k}})), \label{eq: D}
\end{equation}
where $\mathbf{k}$ denotes the crystal momentum with weight $w_\mathbf{k}$. The $i$th band at $\mathbf{k}$ has an eigen energy $\epsilon_{i\mathbf{k}}$, with the Fermi occupation given by $f(\epsilon_{i\mathbf{k}}) = \frac{1}{e^{(\epsilon_{i\mathbf{k}}-\epsilon_\textrm{F})/k_\textrm{B} T} + 1}$. In practice, we also introduce a small smearing $\sigma=0.02$ eV for evaluating $\delta(\epsilon) \approx \frac{1}{\sigma\sqrt{2\pi}} e^{-\frac{x^2}{\sigma^2}}$ due to finite $\mathbf{k}$-mesh used in band structure calculations.

For a fixed band structure, Eq.~\eqref{eq: D} indicates that $D_\textrm{e-h}(\omega, T)$ decreases with increasing temperature due to the Pauli blocking effect. However, this thermal blocking behavior contradicts the experimentally observed increase in $1/q^2$ with temperature. This discrepancy suggests that thermal renormalization of the band structure near the Fermi level plays a significant role. Notably, the band gap in the ZrTe\textsubscript{5} system exhibits strong temperature dependence, undergoing a Lifshitz transition near $T \approx 135$ K, where the band gap closes~\cite{xu2018temperature}. To incorporate this temperature effect in practice, one can adjust the lattice parameters based on experimental measurements~\cite{fjellvaag1986structural}.
In addition, we align the structures such that the band gap(at $\Gamma$) is 30 meV at $T=10$ K, vanishes at $T=135$ K, and becomes 20 meV at $T=300$ K, according to experimental result~\cite{xu2018temperature}, where the biggest change of lattice constant is $b$ from $14.592$ \AA\ to $14.768$ \AA.\ Because the contribution of $D_\textrm{e-h}(\omega_0, T)$ is concentrated at the Brillouin zone center, this allows us to employ an effective uniform $360\times360\times120$ k-grid by only sampling a smaller $20\times20\times20$ sub-grid around $\Gamma$-point for the accurate summation in Eq.~\eqref{eq: D}. 

\begin{figure}[t]
\includegraphics[clip,width=3.5in]{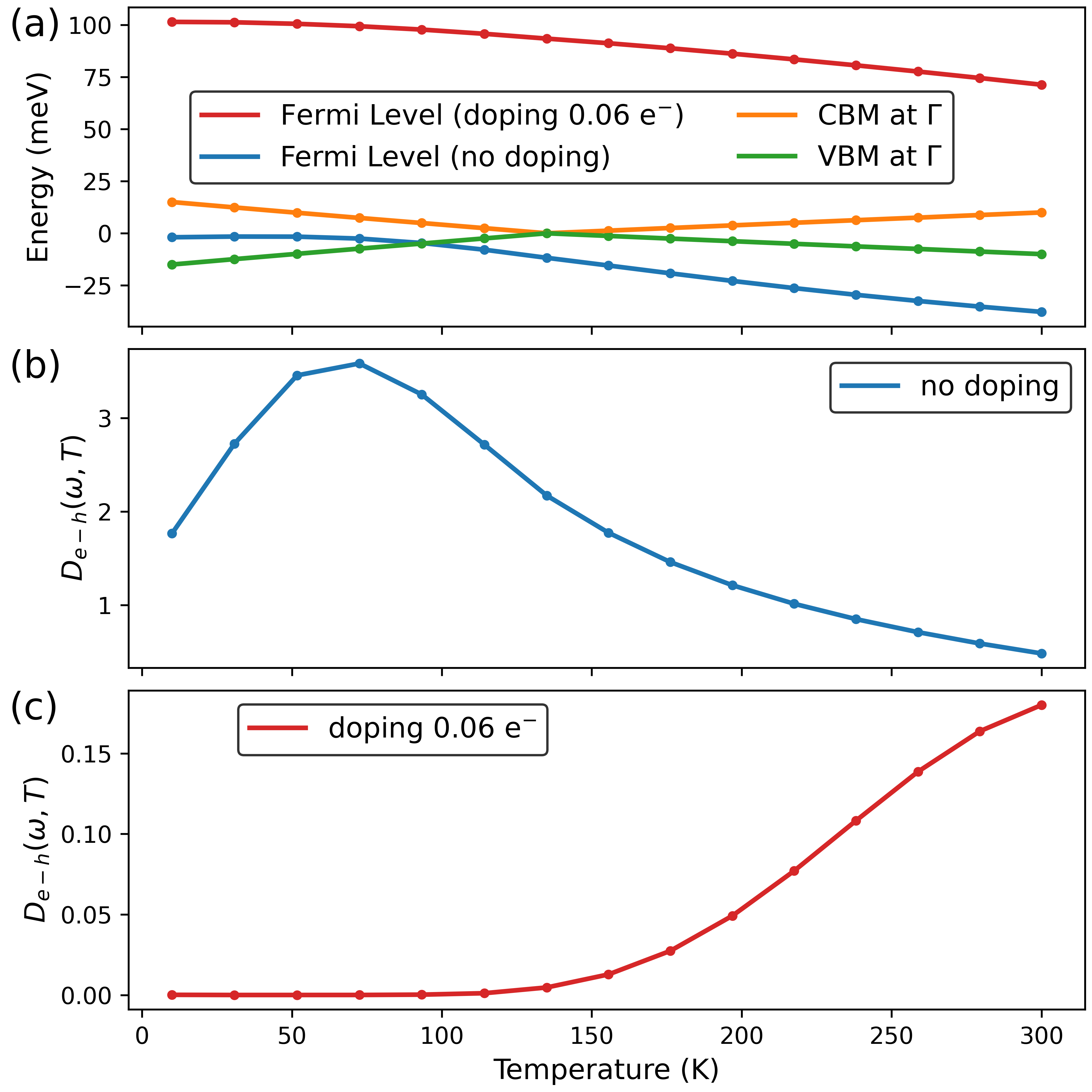}
\caption{
(a) Temperature-dependent Fermi Level without and with electron doping, shown in blue and red curve. Doping electron is 0.06 e$^{-}$. The temperature-dependent VBM and CBM at $\Gamma$ are showing in green and orange curve. (b)Temperature-dependent joint electron–hole pair density of states $D_{e-h}(\omega, T)$ without (b) and with electron doping (c) shown in blue and red curve, respectively.}
\label{Fig4}
\end{figure}

The temperature-dependence of the valence band maximum (VBM) and conduction band minimum (CBM) at $\Gamma$-point is shown in Fig.~\ref{Fig4}(a), together with the Fermi level $\epsilon_\textrm{F}$. In this plot, we shift the middle point between VBM and CBM at zero for clarity. In Fig.~\ref{Fig4}(b) we plot the $D_\textrm{e-h}(\omega_0, T)$ versus temperature in blue solid line. One can see that $D_\textrm{e-h}(\omega_0, T)$ initially increases as temperature rises, reaches a maximum at $T\approx 75$ K, and turns to decrease with further rising temperature. Clearly, the nonmonotonic behaviour of $D_\textrm{e-h}(\omega_0, T)$ is only qualitatively consistent with the experimental results in the low temperature region, where increasing $D_\textrm{e-h}(\omega_0, T)$ implies the enhancement of Fano resonance.

In the above calculations, we assume perfect stoichiometry of an ideal crystal, In experiments, ZrTe\textsubscript{5} samples exhibit some degree of electronic heterogeneity and nanostrip junctions ~\cite{kim2021terahertz}, and excess Te incorporation into the sample during the annealing process can introduce n-type doping~\cite{chi2017lifshitz}. Therefore, we also consider how the temperature dependence of $D_\textrm{e-h}(\omega, T)$ can be affected by small electron doping. Our samples were synthesized using a similar Te-rich flux method as in Ref.~\cite{chi2017lifshitz}, where excess Te was shown to introduce n-type doping. Consistent with that study, our transport measurements revealed negative Hall and Seebeck coefficients at low temperatures, indicating the presence of n-type carriers. Generally, we find that electron doping shifts the peak of $D_\textrm{e-h}(\omega, T)$ toward higher temperature. At doping $0.06$ electron in the primitive cell with Fermi level shifted as shown in Fig.~\ref{Fig4}(a), $D_\textrm{e-h}(\omega, T)$ monotonically grows with temperature $T>100$ K, as shown in the red solid line in Fig.~\ref{Fig4}(c). This analysis implies the electronic heterogeneity in the sample can be an important factor for the experimentally observed Fano resonance enhancement in all the temperature region. We expect that newly developed terahertz nano-imaging methods~\cite{kim2023visualizing,kim2024nano,kim2021terahertz}, especially when applied at low temperatures and in strong magnetic fields\cite{kim2023sub}, will play a pivotal role in pushing forward the ongoing research. 

The above analysis focuses on the key contribution of joint density of states $D_\textrm{e-h}(\omega, T)$ to the temperature-dependence of Fano resonance. However, the electron-phonon coupling strength $V_\mathrm{e-ph}$ in Eq.~\eqref{eq: D} may also depends on temperature and contribute to the variation of $1/q^2$ with temperature. Note that \( V_\mathrm{e-ph} \) can often be estimated from the first derivative of the band energy position or band gap with respect to the phonon eigenmode displacement \( \lambda \)~\cite{dicks2024computationally, yin2013correlation}. However, calculations for the B$_{2g}$ I mode reveal that the band gap exhibits a symmetric dependence on \( \lambda \), with negligible linear terms near equilibrium (\( \lambda = 0 \)). This leads to a vanishing first-order derivative, indicating that a simple linear model cannot capture the coupling strength. Such symmetric behavior is reminiscent of infrared phonon modes in Ta$_2$NiSe$_5$, where the band gap renormalization shows a quadratic dependence on $\lambda$ as described by a two-band model~\cite{michael2024photonic}.

\begin{figure}[t]
\includegraphics[clip,width=3.5in]{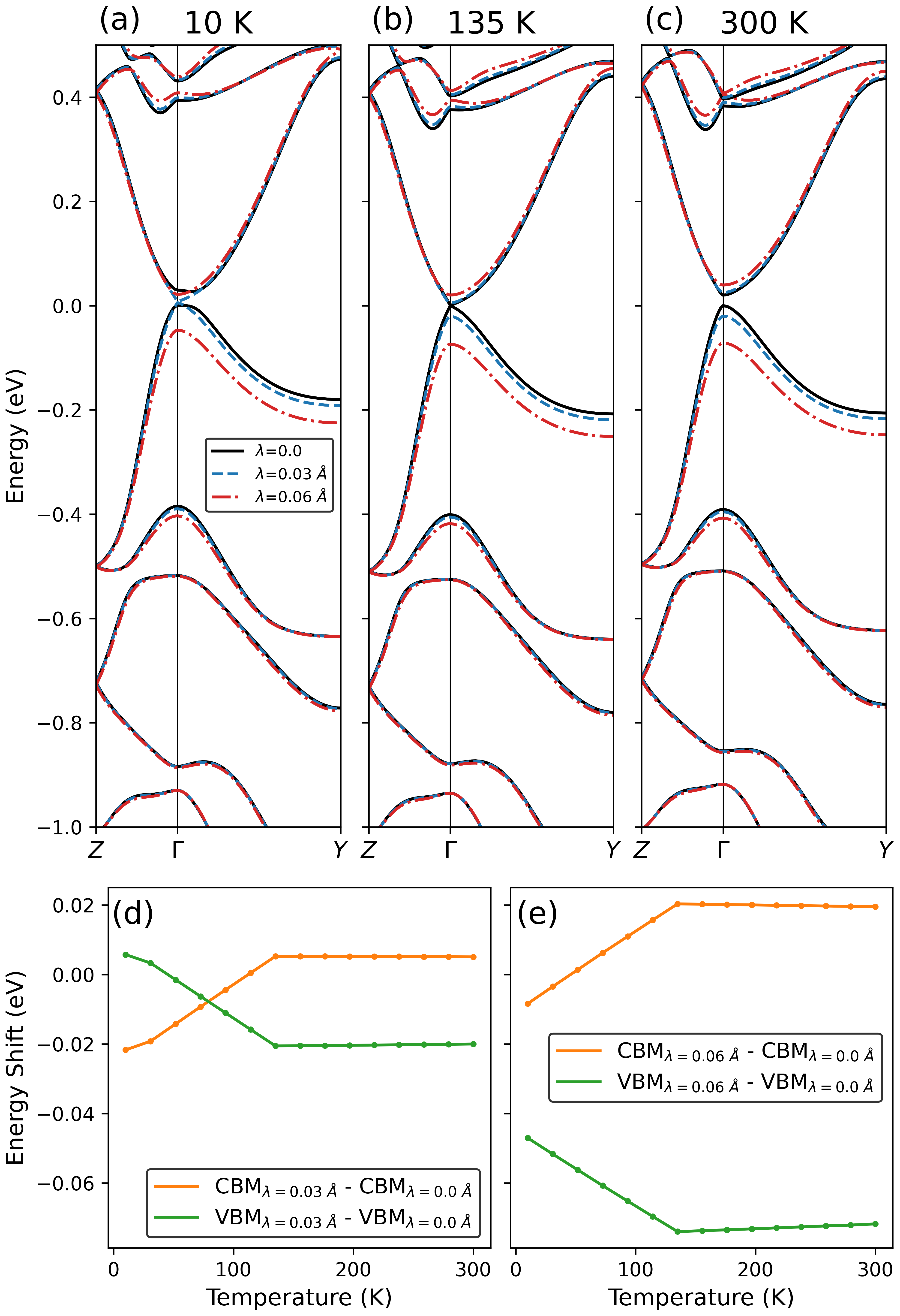}
\caption{
Temperature-dependent band structure shifts induced by B\textsubscript{2g} I phonon mode displacements. Panels (a)–(c) display the electronic band structures at 10 K, 135 K, and 300 K, respectively, for structures subjected to B\textsubscript{2g} \rom{1} phonon distortions. The black solid lines represent the unperturbed (undistorted) structures, while the blue and red dashed lines correspond to displacement amplitudes of $\lambda$ = 0.03 \AA\ and 0.06 \AA,\ respectively. The valence band maximum for $\lambda=0$ is shifted to zero for convenience of comparison across different temperatures. Panels (d) and (e) display the corresponding conduction band minimum (CBM, orange) and valence band maximum (VBM, green) energy shifts for larger amplitudes of $\lambda$ = 0.03 \AA\ and 0.06 \AA, respectively.}

\label{Fig5}
\end{figure}

In contrast, the calculation results for ZrTe$_5$ reveal an asymmetric energy shift between the valence and conduction band edges under B$_{2g}$ I phonon displacements, involving contributions from multiple bands (see Fig.~\ref{Fig5} a-c). This complexity renders a two-band approximation insufficient, and a full multi-band analysis would be required for quantitatively extracting \( V_{\text{e-ph}} \), which we defer to future work. Nevertheless, the overall magnitude of band shifts under finite phonon amplitude remains a meaningful indicator of the electron-phonon coupling strength, which we will use to analyze the trend of \( V_{\text{e-ph}} \) with temperature. 
Fig.~\ref{Fig5} presents the electronic band structures computed at selected temperatures (10 K, 135 K, and 300 K) under B\textsubscript{2g} \rom{1} mode displacements of $\lambda$ = 0.03 \AA\ and 0.06 \AA\ (Te\textsubscript{z} atom displacement). The resulting band shifts, shown in panels (a)–(c), highlight the impact of increasing phonon amplitude on the electronic structure. The black solid lines represent the unperturbed bands ($\lambda=0$), while the blue and red dashed lines correspond to distortions of $\lambda=\pm 0.03$ \AA\ and $\pm 0.06$ \AA, where the band shift is found to be symmetric with respect to $\pm \lambda$. Panels (d) and (e) display the corresponding conduction band minimum (CBM, orange) and valence band maximum (VBM, green) energy shifts for $\lambda$ = 0.03 \AA\ and 0.06 \AA, respectively. The band shifts increase with temperature up to 135 K where the band gap closes, suggesting a temperature-enhancement for the electron-phonon coupling in this regime. Above 135 K, the band shift saturates, indicating a much weaker temperature-dependence of \( V_{\text{e-ph}} \) at higher temperature. This analysis shows that \( V_{\text{e-ph}} \) may contribute to the increase of $1/q^2$ at low temperature regime ($T<135K$), complementary to the contribution of $D_\textrm{e-h}(\omega, T)$ at higher temperatures ($T<135K$) when finite doping is considered.

\setcounter{figure}{0}
\renewcommand{\thefigure}{A\arabic{figure}}

\section{Conclusion}
In this study, the intricate interplay of quantum interference, evident in the Fano resonance, was comprehensively investigated in the context of the Dirac semimetal, ZrTe\textsubscript{5}. The distinct resonance observed in the Raman scattering measurements of ZrTe\textsubscript{5} highlighted the significant difference between two adjacent B\textsubscript{2g} phonon modes. Notably, only the B\textsubscript{2g} \rom{1} mode exhibited the hallmark features of the Fano resonance. Our study found that atomic movements are crucial for understanding the Fano resonance. The size of these movements directly affects the energy gap of the phonon mode. When we looked closer, the B\textsubscript{2g} \rom{1} mode showed a bigger change in its energy gap compared to the B\textsubscript{2g} \rom{2} mode for the same amount of atomic movement. This tells us that electronic structures are especially sensitive to atomic movements of certain phonon modes and how they impact electronic shifts. Data from temperature-based Raman measurements further highlight the importance of atomic movements. Our findings, which showed a growing Fano asymmetry as temperatures rise, reinforce the close relationship between temperature-related atomic changes and the appearance of the Fano resonance.

Moreover, first-principles calculations provided an invaluable perspective into the inherent electronic states continuum associated with the B\textsubscript{2g} \rom{1} mode and the salient absence of the Fano resonance in the B\textsubscript{2g} \rom{2} mode. Especially, the eigenvector orientation concerning the Te\textsubscript{z} and Te\textsubscript{d} atoms in the two B\textsubscript{2g} modes emerged as a significant factor in determining their susceptibility to Fano resonance.

This study has not only deepened our understanding of the interplay between atomic movements and Fano resonance in Dirac semimetals but has also unveiled intriguing insights that may be pivotal in shaping future research and applications in high-speed electronics, spintronics, and quantum computing.

\section{Acknowledgements}
The THz spectroscopy measurement was supported by the US Department of Energy, Office of Basic Energy Science, Division of Materials Sciences and Engineering (Ames National Laboratory is operated for the US Department of Energy by Iowa State University under contract no. DE-AC02-07CH11358). This work was supported also by the grant of computer time at the National Energy Research Scientific Computing Center (NERSC) in Berkeley, California.

\bibliography{refabbrev, ref}

\end{document}